# Premelting-directed interfacial design of nanoshell-coated cells in directional freezing and thawing

*Jiaxue You[1,2], Yunhan Zhang[1], Qiao Chen[1], Yifan Xie[1], Yixiao Li[1], Lifang Hu[3] and Li Shang[1,4]\**

[1] State Key Laboratory of Solidification Processing, Center for Nano Energy Materials, School of Materials Science and Engineering, Northwestern Polytechnical University, Xi'an, 710072, China

[2] Key Laboratory of Applied Surface and Colloid Chemistry, National Ministry of Education, Shaanxi Key Laboratory for Advanced Energy Devices, Shaanxi Engineering Lab for Advanced Energy Technology, Institute for Advanced Energy Materials, School of Materials Science and Engineering, Shaanxi Normal University, Xi'an 710119, China

[3] Key Laboratory for Space Bioscience and Biotechnology, Institute of Special Environmental Biophysics, School of Life Sciences, Northwestern Polytechnical University, Xi'an 710072, China

[4] NPU-QMUL Joint Research Institute of Advanced Materials and Structures (JRI-AMAS), Northwestern Polytechnical University,
Xi'an, 710072, China

*Corresponding author: Li Shang, E-mail: li.shang@nwpu.edu.cn

**Abstract**

Freezing interface interacting with soft cells is a core issue in cryopreservation. However, most of existing cryopreservation approaches face challenges such as complex processing and poor controllability. Herein we report a new, nanotechnology-based method to modulate interfacial interactions between ice and cells *via* the premelting theory. Through the interfacial design by controllably modifying biocompatible nanoshells on the surface of cells, the effective Hamaker constant between cells and ice can be modified. The thickness of premelted films between cells and ice is further regulated, and directional migration of coated cells occurs in a thermal gradient to achieve directional thawing of coated cells. The decreased mortality of freezing coated cells suggests the premelting modulation in freezing and thawing can effectively protect cells from mechanical damage of ice formation during freezing and ice recrystallization during thawing.

**Keywords**: freezing, interfacial design, nanoshells, cell/ice interactions, premelting

# 1. Introduction

Freezing and thawing, a classical liquid/solid phase transformation, is widely involved in many fields, such as alloy solidification and smelting, frozen soils, food science and biological cryopreservation[1-7]. In many cases, freezing interfaces encounter hard or soft objects such as inorganic particles, gas bubbles, liquid droplets, or biological cells, for example, mammalian cells will be physically damaged when engulfed by growing ice crystals[8]. The formation and growth of ice are highly unfavorable in most biological systems. Ice formation and growth not only lead to mechanical damage on a cellular level, but also osmotic shock as the concentration of extracellular solutes rises as the liquid water volume fraction decreases.[8] Moreover, ice recrystallization during the thawing process also causes cell death because the migrating grain boundaries of ice shear and deform cells.

The popular solution in freezing cells is using cryoprotectants, such as DMSO (dimethyl sulfoxide), which are intrinsically toxic.[9] Another method is using polymers or proteins to modify the ice crystal growth through the curvature effects.[10, 11] Also the latest cryoprotectant-free technology has been developed for freezing and thawing of cells using liquid marbles filled with hydrogel.[12] However, the intrinsic crystallographic behavior of ice is complex[10] and hard to be modulated *via* these strategies. It has been reported that failures in effective cryopreservation result in the loss of nearly half of all global biological objects such as vaccines.[5] To this end, it is of great importance to develop new, efficient cryopreservation strategies.

While the conventional cryopreservation method is to change the process of freezing ice, an alternative approach is to modify the surface of soft cells in order to protect them from mechanical injuries at the ice/cell interface. The core issue in cryopreservation is interfacial interactions between ice and biological objects.[1, 13] The premelting theory has well revealed interfacial interactions between ice and foreign substrates.[1, 14, 15] When the ice meets a foreign substrate, a premelting liquid layer with a thickness of several nanometers will appear between the ice and the substrate due to the interfacial energy difference between ice, the liquid layer and the substrate.[14] This premelting liquid layer can buffer cells against mechanical injuries of ice growth. Moreover, this layer can cause the film flow and drive the substrate moving to warm side when a thermal gradient is applied, which is called as thermal regelation.[16] According to the premelting theory and thermal regelation, it is possible to modify the surface of cells and regulate the thickness of premelting layer to achieve the directional thawing of cells which can avoid the migrating grain boundaries of ice shearing cells.

Recent advances in micro- and nano-encapsulation of cells provide efficient strategies to modify the surface of cells, which can transfer soft cells into hard particles.[17-24] Many different kinds of biocompatible nanoshells, e.g. silica[25], titania[26], graphene[27], polydopamine[28], manganese dioxide[29] or tannic acid (TA)-$Fe^{3+}$ complexes[30-33], have been fabricated on the surface of microbial and mammalian cells. Generally, these nanoshells can be divided into two categories according to their degradability. The nanoshells, such as silica, are hard to be degraded, while the nanoshells of TA-$Fe^{3+}$ complexes are easily degradable under cytocompatible conditions by adding few dilute hydrogen chloride (HCl). Different shells cause different strengths of interactions between cells and ice as well as different thicknesses of liquid layers according to the premelting theory.

In this work, we modified the surfaces of cells with different nanoshells and then directionally thawed them in a self-designed thermal gradient apparatus after freezing, as illustrated in **Scheme 1**. The directional thawing of coated cells was in-situ observed in a thermal gradient, while there was almost no migration of native cells in the ice. This distinct behavior was further analyzed by the premelting theory. Finally, by the cell viability test, we showed that premelting-directed interfacial modification improved viability after freezing and thawing.

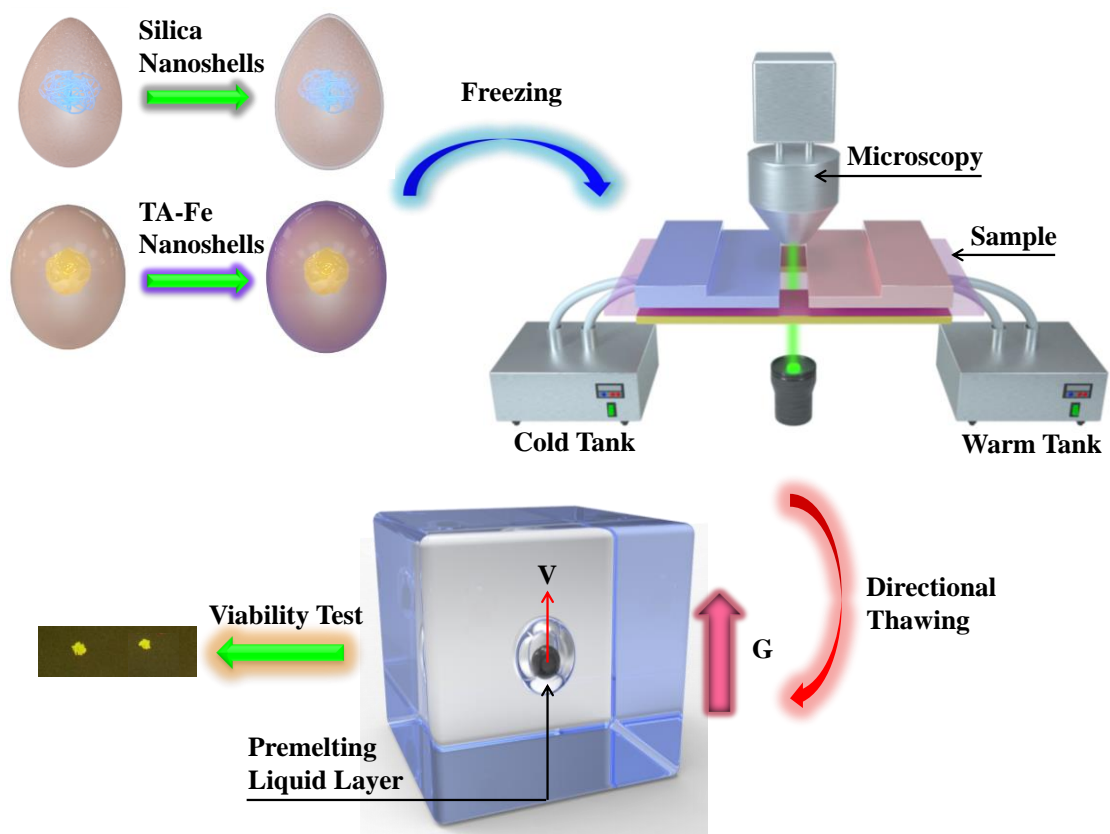

**Scheme 1** Procedure for cell encapsulation and freezing-thawing in a thermal gradient. The coated cells are frozen in the ice and put into a thermal gradient apparatus for directional thawing. Viability of coated cells after thawing was then checked by confocal fluorescence microscopy.

## 2. Experimental methods

*Cell Culture*: Mammalian cells (1 ml, $10^3$ per ml) were added to the culture dish which contained Dulbecco's Modified Eagle Medium (DMEM, 4 ml). The cultivation conditions were 37 $^0$C, 5% $CO_2$ concentration, and sterile. After two days, trypsin (1 ml) and phosphate buffer saline solution (PBS, sterile, 1 ml) were added to the culture dish to digest cells for 5 minutes. Then the cells were suspended in the buffer solution. The suspension was centrifuged for 3 minutes at 1500 r/min, and the supernatant was discarded. The centrifuged cells were washed three times by sterile PBS solution (5 ml).

Dry yeast cells (10 mg) were cultured in the yeast-extract-glucose liquid media with continuous shaking at 30 ℃ for 12 h. Then, yeasts were harvested, washed with deionized (DI) water three times and dispersed in DI water ($OD_{600}$=1.8, $OD_{600}$: the optical density at the wavelength of 600 nm). Optical density was measured with a microplate reader (SynergyTMHT Ser, USA).

*Coat Cells with Nanoshells*: TA-Fe$^{3+}$ nanoshells were fabricated based on the method reported in the literature.[30, 31] Aqueous solution of TA (5 μL, 40 mg/mL) and FeCl$_3$·6H$_2$O (5 μL, 10 mg/mL) were added sequentially to the suspension of cells (490 μL) with vigorous mixing for 10 s. After adding FeCl$_3$·6H$_2$O solution, the resulting suspension was mixed vigorously for 10 s, and 3-(N-morpholino) propanesulfonic acid (MOPS) buffer solution (0.5 mL, 20 mM, pH 7.4) was added to the cells@[TA-Fe$^{3+}$] suspension for stabilizing the pH, resulting in the formation of stable TA-Fe$^{3+}$ shell. Cells@[TA-Fe$^{3+}$] were washed with sterile PBS solution three times to remove any residual TA and FeCl$_3$. The coating process was repeated four times ([TA-Fe$^{3+}$]$_4$) to form nanoshells.

Silica nanoshells were fabricated based on the method reported by Ref.[34]. The cationic poly(diallyldimethylammonium chloride) (PDDA) and the anionic sodium polystyrene sulfonate (PSS) were alternately deposited onto the surface of the yeast cells. Briefly, the yeast cells were immersed alternately in solutions of PDDA (5 mg/mL) and PSS (5 mg/mL) for 5 min each, which led to the coating of individual yeast cells with PDDA/PSS multilayers (7/6) (7/6 means 7 layers of PDDA and 6 layers of PSS). Then 50 mM silicic acid solution containing the PDDA/PSS multilayer-coated yeast was stirred at room temperature for 30 min, leading to formation of silica-encapsulated yeast (yeast@SiO$_2$). After 30 min, the substrate was removed and washed with 0.5 M aqueous NaCl solution.

*Characterization of Coated Cells*: We explored the morphology and elemental distribution of nanoshells *via* optical microscope, scanning electron microscope (SEM), energy dispersive spectroscopy (EDS), Raman spectroscopy and X-ray photoelectron spectroscopy (XPS). Field-emission scanning electron microscopy (FE-SEM, FEI Verios G4) imaging was performed at an accelerating voltage of 18 kV. Raman spectra were measured by using a Raman microscope (Renishaw, UK) with 532 nm laser light as excitation source, and the laser power was set at 0.5 W. XPS measurements were carried out on a K-Alpha XPS spectrometer (Kratos Analytical, Japan), using Al Kα X-ray radiation (1486.6 eV), Slot collimation mode and emission current 10 mA for excitation.

*Freezing and Directional Thawing*: The suspensions of nanoshell-coated cells were added into rectangular glass capillary tubes [35, 36] and frozen in the refrigerator (-20 $^0$C) for 14 hours or 72 hours. Then the frozen suspensions were put into a self-designed apparatus of thermal gradient for directional thawing, as shown in Scheme 1.[35, 36] In the apparatus, the thermal gradient was produced by a heating zone and a cooling zone separated by a gap of 5 mm. In-situ observation was achieved through an optical microscope stage (ZEISS Axio Scope.A1) with a charge-coupled device (CCD) camera. The thermal gradient was measured as G=9.7 K/cm. The cold side was aound -10 $^0$C and the warm side was around 7 $^0$C. Heat loss caused by the ambient temperature occured during the thermal cycling. Images were recorded *via* a CCD camera with 2580×1944 sensitive elements on a time-lapse video recorder and further analyzed by the software image processing (Image Pro plus 6.0).

*Cell Viability Test*: Cell viability of the native and coated cells was assessed with live/dead assay based on propidium iodide (PI) and fluorescein diacetate (FDA) staining.[37] FDA is hydrolyzed to the green-fluorescent fluorescein by esterases in the metabolically active cells, while the red fluorescent molecule PI can enter only nonviable cells to stain nucleic acids of dead cells. The FDA stock solution (5 mg/mL) was prepared in DMSO due to the insolubility of FDA in water. The PI stock solution (2 mg/mL) was prepared in mannitol. Stock solutions of FDA (10 μL) and PI (8 μL) were mixed with the cell suspension (0.5 mL). After incubation for 20 min at room temperature, the cells were washed with sterile PBS solution three times and then visualized by confocal laser-scanning microscopy (Leica SP8). The excitation wavelengths of fluorescein diacetate (FDA) and propidium iodide (PI) are 494 nm and 537 nm, respectively. The emission wavelengths of FDA and PI are 520 nm and 617 nm, respectively.

*Materials*: Tannic acid (TA, Sigma), iron(II) chloride hydrate (Sigma), iron(III) chloride hexahydrate ($FeCl_3·6H_2O$, Sigma), 3-(N-morpholino)propanesulfonic acid (MOPS, Sigma), fluorescein diacetate (FDA, Solarbio), propidium iodide (PI, Solarbio), hydrochloric acid (HCl, 37 wt%, Hushi), phosphate-buffered saline (PBS, pH 7.4, Genview), Dulbecco's modified eagle medium (DMEM, GIBCO) and trypsin (GIBCO) were used as received. Poly(diallyldimethylammonium chloride) (PDDA, Mw: 100000–200000, 20 wt% in $H_2O$, Sigma) and sodium polystyrene sulfonate (PSS, average Mw: 70000, powder, Sigma) solutions, and tetramethyl orthosilicate (TMOS, 100 mM, Wokai Company, CN) were prepared. Yeast-extract-glucose liquid media was prepared with 100 mg of yeast extract and 300 mg of glucose (Sigma) in 7 mL of deionized (DI) water and used after autoclaved (15 min, 121 °C). Ultrapure water (18.25 MΩ·cm, the Molecular Company) was used.

## 3. Results and Discussions

### 3.1 Cell Encapsulation

Cells were readily coated with the coordination-based shells through sequential addition of TA and $Fe^{3+}$ to cell suspensions. Each iron ion center could bind three galloyl groups of TA, resulting in the cross-linking of TA and $Fe^{3+}$ on the cell surface.[30] As shown in Figure 1, upon addition of TA and $Fe^{3+}$, the cell suspension immediately turned purple. The opacity change was attributed to the formation of TA–$Fe^{3+}$ complexes on the surface of cells.

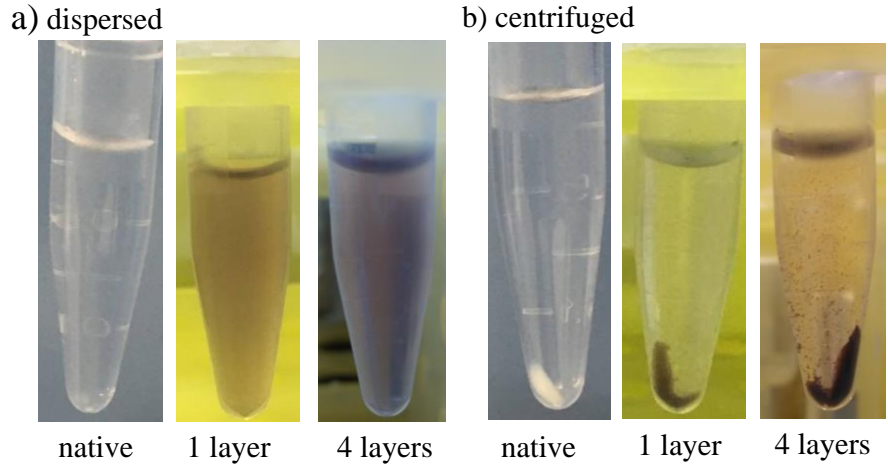

a) dispersed     b) centrifuged

native   1 layer   4 layers     native   1 layer   4 layers

**Figure 1** Macro observations of native bone cells and tannic acid (TA)-$Fe^{3+}$ coated cells suspended in the PBS solution before (a) and after (b) the centrifugation. The native cells are white while the coated cells are purple.

The shell-coated cells were then characterized by scanning electron microscopy (SEM). As shown in **Figure 2**a, the surface of murine bone cells@[TA–$Fe^{3+}$]$_4$ was rather rough. According to previous studies, the average thickness of the nanoshells is about 10 nm, when a TA–$Fe^{3+}$ layer was sequentially deposited on the surface of cells.[32] The energy dispersive spectroscopy (EDS) measurement further confirmed the presence of Fe element in the shell (Figure 2b). Successful shell formation was also supported by the Raman spectrum of bone cells@[TA–$Fe^{3+}$]$_4$ (Figure 2c). The peaks at around 590, 1354, and 1491 cm$^{-1}$ are attributed specifically to the chelated $Fe^{3+}$ by the oxygen atoms of the catechol.[38] X-ray photoelectron spectroscopy (XPS) measurements were carried out to analyze the valence states of iron in the bone cells@[TA–$Fe^{3+}$]. The Fe 2p XPS spectrum (Figure 2h) shows the binding energy (BE) of Fe $2p_{1/2}$ and Fe $2p_{3/2}$ at 725.3 eV and 712.0 eV, respectively. The BE value of Fe $2p_{3/2}$ (712.0 eV) for the nano-films is similar to the BE value of Fe $2p_{3/2}$ (712.4 eV) for $FeCl_3$,[39, 40] which confirms the existence of $Fe^{3+}$. On the other hand, the TA–$Fe^{3+}$ shell can be degraded by adding HCl [30]. The solution color of cells@[TA–$Fe^{3+}$]$_4$ will turn to white when HCl is added, indicating the shell degradation. After 30 or 90 min treatment with 20 mM HCl, the cells after shell degradation dramatically increase similar to the proliferation of native cells, confirming the cytocompatible, mild conditions for shell degradation in previous studies.

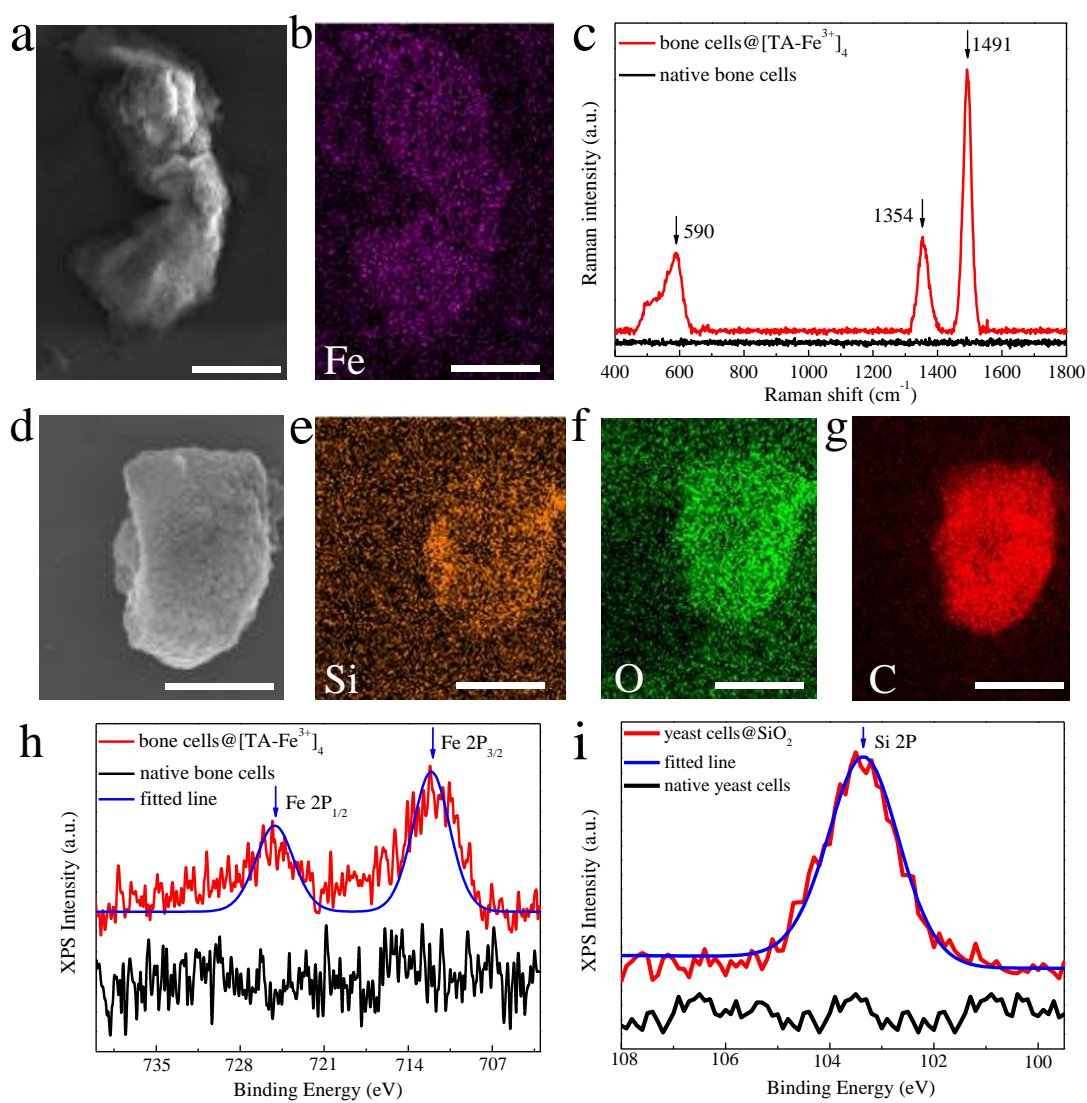

**Figure 2** Nanoshells characterization. (a) SEM images of bone cells@[TA-$Fe^{3+}$]$_4$. (b) EDS spectroscopy plane profiles of Fe for coated bone cell shown in (a). The scale bars are 25 μm. (c) Raman spectra of native bone cells (black) and cells@[TA-$Fe^{3+}$]$_4$ (red). Black arrows indicate the strong bands attributed to the ring structures of TA. (d) SEM image of a yeast cell@$SiO_2$. The scale bar is 2.5 μm. (e-g) EDS spectroscopy plane profiles of Si, O, C for coated yeast cells. The scale bars are 2.5 μm. (h) X-ray photoelectron spectroscopy (XPS) analysis of native bone cells (black), and cells@[TA-$Fe^{3+}$]$_4$ (red). Blue arrows indicate the valence state of Fe. (i) XPS analysis of native yeast cells (black), and yeast@$SiO_2$ (red). Blue arrows indicate the valence state of Si.

Yeast cells covered by silica nanoshells were also characterized by SEM and XPS in Figure 2d and 2i, respectively. According to previous studies, the average shell thickness of the nanoshells was increased by 5 nm,[34] when a PDDA/PSS layer was sequentially deposited on cells. PDDA and PSS are precursors reacting with the silicic acid solution to form silica nanoshells. SEM images showed that the surface of yeast cell@$SiO_2$ is much rough (Figure 2d), while the surface of native yeast cells is smooth.[29] The results of EDS in Figure 2e and Figure S1 indicate the presence of Si element on the nanoshell. The Si 2p XPS spectrum (Figure 2i) shows the BE of Si 2p at 103.3 eV, which is similar to the standard BE values of $SiO_2$ at 103 eV and standard BE of $SiO_2$–$nH_2O$ at 103.5 eV.[41] This result further confirmed the presence of $SiO_2$ and $SiO_2$–$nH_2O$ on the surface of yeast cells. The yeast cells can also be covered by nanoshells of TA-$Fe^{3+}$, as confirmed by Raman spectra (Figure S2).

**3.2 Directional Thawing of Nanoshell-coated Cells**

Upon successfully modifying the cell surface with nanoshels, we next investigated their directional thawing behavior. The cells frozen in ice were placed in a thermal gradient apparatus. Upon applying a thermal gradient, we can observe remarkable behavior differences between native cells and coated cells. Cells coated with nanoshells can migrate from the cold side to the warm side and finally move out of the ice, while native cells stay still in the ice without any observable movement. **Figure 3**b and Figure S3 show the migration of bone cells@[TA-$Fe^{3+}$]$_4$ in ice in two minutes, while no movement was observable for native mammalian cells even after ten minutes. We also observed similar behaviors of yeast@$SiO_2$ and native yeast in the ice, as shown in Figure 3c and 3d. The migration of yeast@$SiO_2$ in ice is strong while native yeast cells show very weak migration. The position of 0 °C is defined as zero and minus position means minus teperature. The position-time curves of coated cells and native cells (**Figure 4**a-b) showed the migration speed of coated cells increases with time while that of native cells stays zero.

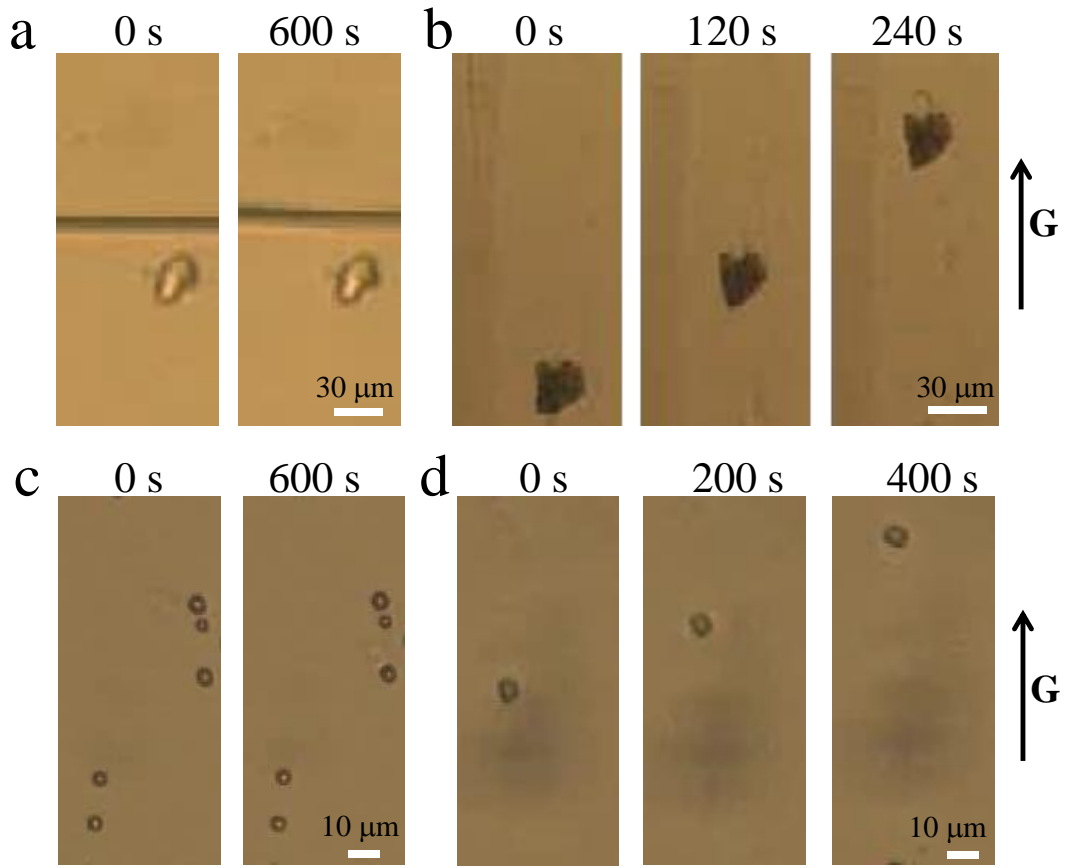

**Figure 3** In-situ observations of thermal regulation of native mammalian cells (a) and coated mammalian cells (b) in the ice under a linear thermal gradient G. In-situ observations of thermal regulation of native yeast cells (c) and coated yeast cells (d).

The interaction mechanism between cells and ice can be well described by the premelting theory and thermal regulation.[42] When a cell coated by nanoshells is embedded in ice that is close to its bulk melting temperature (Figure 4d), there will be a premelted thin film between the nanoshell and ice. An imposed temperature gradient induces coated cells to move by a process of melting and refreezing. The net thermo-molecular force pushes coated cells to warm side.[16, 43] On the other hand, the liquid viscous force drags coated cells to the cold side.

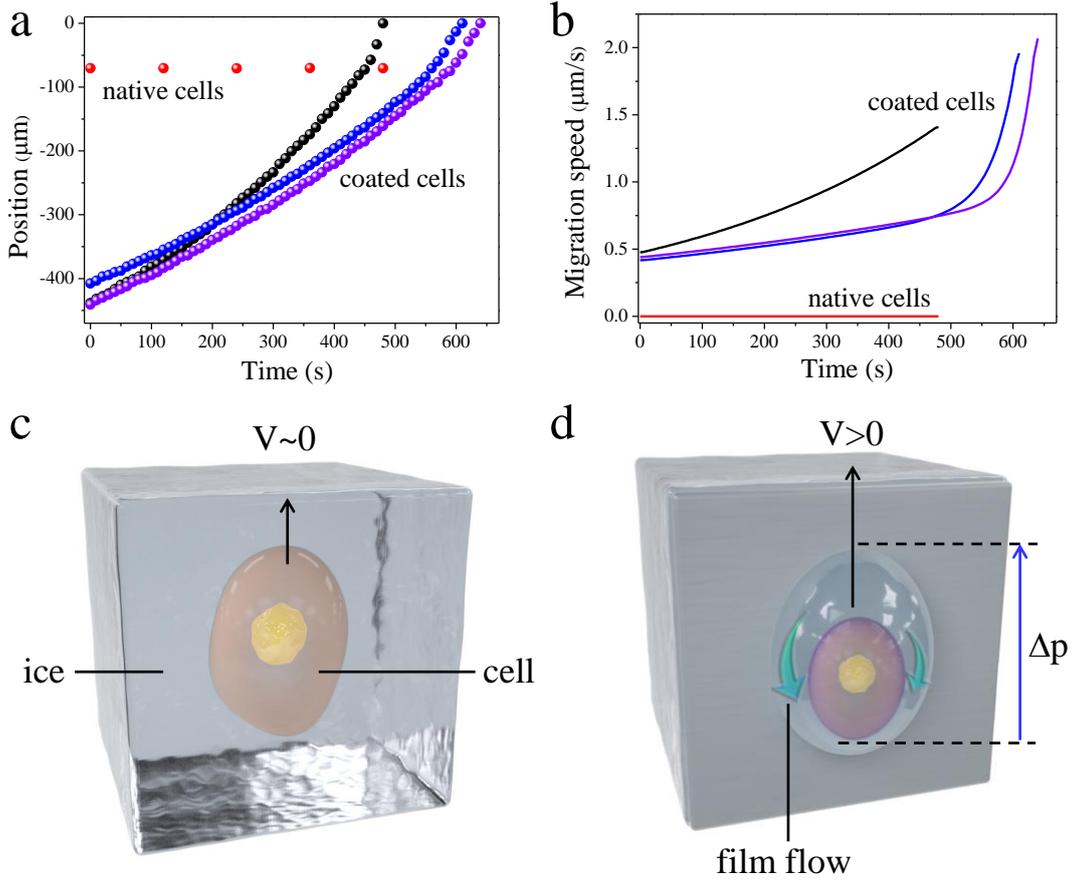

**Figure 4** (a) Position-time curves for coated cells and native cells. (b) Migration speed-time curves for coated cells and native cells. (c-d) Schematics of thermal regelation for native mammalian cells (c) and coated mammalian cells (d) in the ice under a linear thermal gradient G. Coated cells are protected by the premelting liquid film. The arrows are liquid flow from high temperature to low temperature, which is driven by pressure difference Δp.

The net thermo-molecular force on the coated cells (Figure 4), arising from the intermolecular interactions[42], is

$$F_T = \int_s p_T ds = \int_s [\frac{\rho_s L}{T_m}(T_m - T_i) - \gamma_{sl} k] ds = \frac{m_s L}{T_m} G, \qquad (1)$$

where $p_T$ is disjoining pressure, the surface area element $ds$ points in the direction of the outward normal to the coated cells. $\gamma_{sl}$ is solid/liquid interface energy, $k$ is the interface curvature around coated cells. $\rho_s$ is the density of ice. $L$ is the latent heat. $T_m$ is the equilibrium freezing point. $T_i$ is the temperature at the center of the coated cells. $m_s = \rho_s V_p$ is the mass of displaced ice. $V_p = 4\pi a^3 / 3$ is the volume of the cell. $a$ is the radius of cells. $G$

is the imposed thermal gradient. On the other hand, lubrication theory gives volumetric flow rate

$$q = -\frac{d^3}{12\mu a}\frac{\partial p_l}{\partial \theta}, \qquad (2)$$

where $d$ is the thickness of premelted thin film, $\mu$ is the dynamical viscosity of water, $p_l$ is the pressure in the premelted thin film, $\theta$ is the polar angle measured from the downward vertical. Mass conservation gives

$$-2\pi a q \sin\theta = \pi a^2 V \sin^2\theta, \qquad (3)$$

where $V$ is the refreezing speed of ice (or the migration speed of the coated cells). Combine Eqs.(2) and (3) to find the liquid pressure

$$p_l = p_0 + \frac{6\mu a^2 \cos\theta}{d^3}V, \qquad (4)$$

where $p_0$ is the external pressure. Therefore, liquid viscous force $F_f$ can be calculated as

$$F_f = -\int_s p_l ds = -\frac{4\pi a^3}{3}\frac{6\mu a}{d^3}V, \qquad (5)$$

where $d = \lambda(A) \times (\frac{T_m}{T_m - T_i})^{1/\upsilon}$ is the premelting film thickness. The constants $\lambda(A)$ and $\upsilon$ are physical parameters determined by the type and range of interactions between cells and ice. $A$ is the effective Hamaker constant. The $T_i$ in a linear thermal gradient can be approximated as

$$T_i = T_m + Gz_i, \qquad (6)$$

where $z_i$ is the center position of the coated cells. Force balance on the coated cells requires

$$F_T + F_f = 0. \qquad (7)$$

Hence the migration speed of coated cells is

$$V = \frac{\rho_s L G}{T_m}\frac{d^3}{6\mu a}, \qquad (8)$$

When coated cells move to the warm side, temperature of coated cells $T_i$ increases, leading to the increase of premelting film thickness $d$. According to **Equation 8**, the migration speed of coated cells increases with the increase of premelting film thickness. Therefore, the migration speed of coated cells increases with time, which is consistent with the experimental results in Figure 4b.

The mechanism of stilled native cells in ice can also be well described by Equation 8. When native cells contact with ice, the effective Hamaker constant ($A_n$) is much smaller than that between coated cells and ice ($A_c$). Therefore, the thickness of premelted thin film ($d_n$) between native cells and ice is much smaller than that between coated cells and ice ($d_c$). Accordingly, the migration speed of native cells in the thermal gradient is almost zero according to Equation 8. The changed effective Hamaker constant of coated cells can be confirmed by Figure 5.

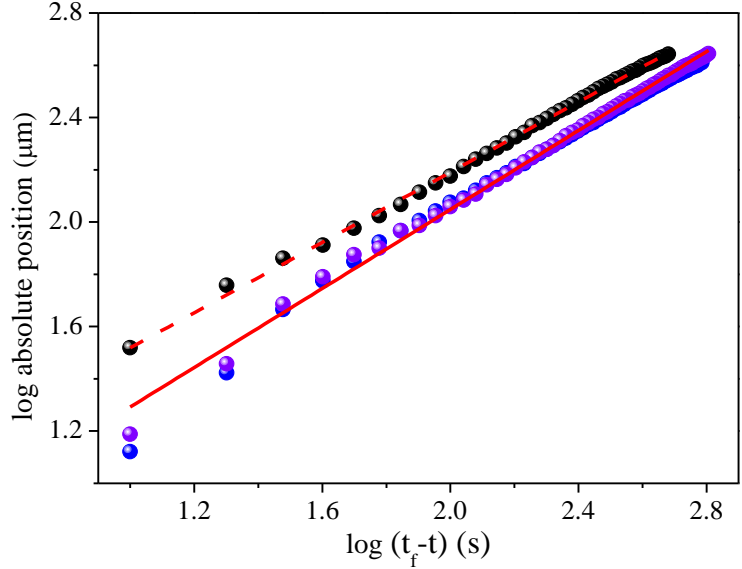

**Figure 5** Log scales of position versus time, fitted by the red lines. $t_f$ is the time that coated cells finish travel in ice.

Figure 5 is the log plot of Figure 4 a. From Figure 5, the position can be calculated as [44]

$$\Delta x = m(t_f - t)^n, \tag{9}$$

where $\Delta x$ is the absolute position. $t_f$ is the time that coated cells finish travel in ice. The fitting parameters $m$ and $n$ can be fitted from Figure 5. By differentiation, $V$ can be calculated as

$$V = \frac{\partial \Delta x}{\partial t} = \frac{n \Delta x}{t_f - t}. \tag{10}$$

The velocity as a function of the undercooling can be established as [44]

$$V = K \Delta x^p, \tag{11}$$

where $K = \frac{\rho_s L}{6 \mu a} \lambda^3 (\frac{G}{T_m})^{1-3/\upsilon}$ and $p = -3/\upsilon$. Upon integration, Equation (11) becomes

$$\Delta x = [K(1-p)(t_f - t)]^{1/(1-p)}. \tag{12}$$

Therefore, $K = mn$. From the fitting of Figure 5, m=5.1 and n=0.7. $\lambda$ is around 1 μm. The calculation of $\lambda$ from TA-$Fe^{3+}$ complexes is three orders of magnitude larger than the common value in Ref.[44]. $\lambda(A)$ is proportional to the effective Hamaker constant $A$. Therefore, the effective Hamaker constant of cells increased after coating. Moreover, the changed effective Hamaker constant of coated cells can also be confirmed by the directional freezing experiments in Figure 6.

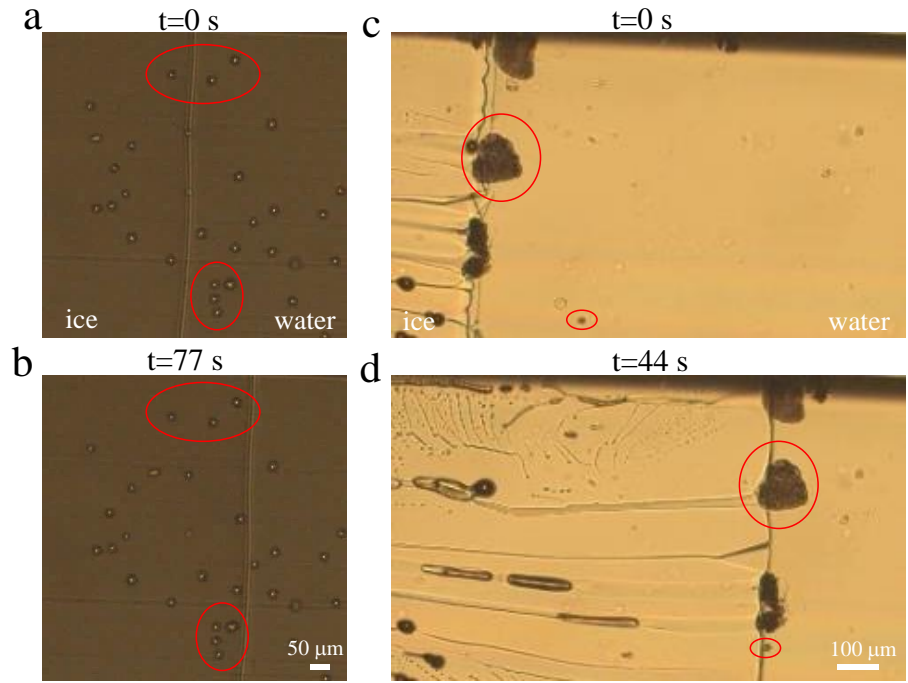

**Figure 6** In-situ observations of directional freezing native bone cells and coated bone cells. (a-b) native cells are engulfed by the freezing interface marked in the red circles under a pulling speed 2.5 μm/s. (c-d) coated cells are pushed away by the freezing interface marked in the red circles under a pulling speed of 16.7 μm/s or of 60 μm/s.

We directionally froze native cells and coated cells under the same condition using the thermal gradient apparatus. When a pulling speed of 2.5 μm/s was applied, native cells were engulfed by the freezing interface, while coated cells were pushed away under a pulling speed of 16.7 μm/s or of 60 μm/s. It means the critical engulfed speed of native cells is much smaller than that of coated cells. According to the single particle model (**Equation 13**),[45, 46] a decreased effective Hamaker constant decreases the critical engulfed speed $V_c$. Therefore, the engulfed speeds in the directional freezing experiments indicate the effective Hamaker constant of native cells ($A_n$) is much smaller than that of coated cells ($A_c$). The difference between them is bigger than two orders of magnitude ($A_c > 100\ A_n$) according to Equation 13.

$$V_c \propto \frac{(\rho_S LG)^{1/4} A^{3/4}}{\mu T_m^{1/4} a}. \tag{13}$$

Figure 4d shows that through the interfacial design between cells and ice at the nanoscale, i.e., adding biocompatible and degradable nanoshells on the surface of cells, the effective Hamaker constant $A$ between cells and ice increases. According to Equation 8, the thickness of premelted film between cells and ice is further increased, and directional migration of coated cells occurs at macro-scale in a thermal gradient to achieve directional thawing of coated cells. The directional migration surrounded by premelted films protects cells from shear forces of migrating grain boundaries in ice recrystallization during the thawing process. Therefore, mortality of coated cells is expected to be decreased after freezing and thawing. Note that since the thermal gradient is from -10 $^o$C to 7$^o$C, there is an interfacial line between ice and liquid that the cells have to pass and the migration in liquid ceases.

**3.3 Cell Viability Test**

We then investigated the viability of these encapsulated cells *via* FDA assay, which was used to examine the integrity of cell membrane and cellular esterase activity. As illustrated in **Figure 7**, the viability of coated cells and native cells were tested. Cell viability statistics is based on the result of more than three hundred cells. The membrane integrity of coated cells was confirmed in previous works [31]. A fluorophore-conjugated protein (BSA-Alexa Fluor 647) was immobilized onto coated cells, and the core–shell structure was observed for viable coated cells.

The initial viability of native cells is around 92%. After coating the initial viability decreases to 72%. The viability of the coated bone cells after freezing and thawing is over 27%, which is higher than that of native cells, 17%. Similar result was also observed in the previous study of yeast cell encapsulation, where the viability of coated yeast cells and native yeast cells is 56% and 24%, respectively.[34] Apparently, in both cases, the viability of cells after the freezing/thawing processing is improved when coating with nanoshells. We note that the cell viability of our coated cells is lower than previous reports,[30, 34] which is possibly because mammalian cells and a lower temperature of cryopreservation (-20 $^{o}$C) were used in the present study. Note that the initial mortality of coated cells caused by the nanoshells can be decreased by careful operations. In previous reports, the viability of the coated cells was about 95%, which was similar to that of native cells. The introduced materials of nanoshells have no harm to cells. Therefore, the cell mortalities between coated cells and native cells in freezing and thawing can be used to reveal the protection of nanoshells from ice. The mortality of native cells in freezing and thawing, 75%, is much higher than that of coated cells 45%. The viability tests of freezing coated cells for 72 hours and freezing cells with DMSO for 14 hours were also presented in Figures S4 and S5, respectively. After 72 hours freezing, the viability of coated cells is still higher than that of native cells. After thawing, the TA–$Fe^{3+}$ shell can be degraded by adding extreme-dilute HCl in cytocompatible, mild conditions in previous studies. The decrease in the mortality suggests that our premelting modulation in freezing and thawing can effectively protect coated cells from mechanical damage of ice formation during freezing and ice recrystallization during thawing.

The mortality of coated cells in freezing and thawing should be attributed to the inner ice growth of cells in the freezing process. Nanoshells can only protect the extrinsic ice growth but have no help on the inner ice growth. Moreover, the nanoshell of TA-$Fe^{3+}$ complexes is not rigid enough to fully resist the extrinsic ice growth. If a much rigid nanoshell is used such as carbon diamond, the viability of coated cells may be further improved.

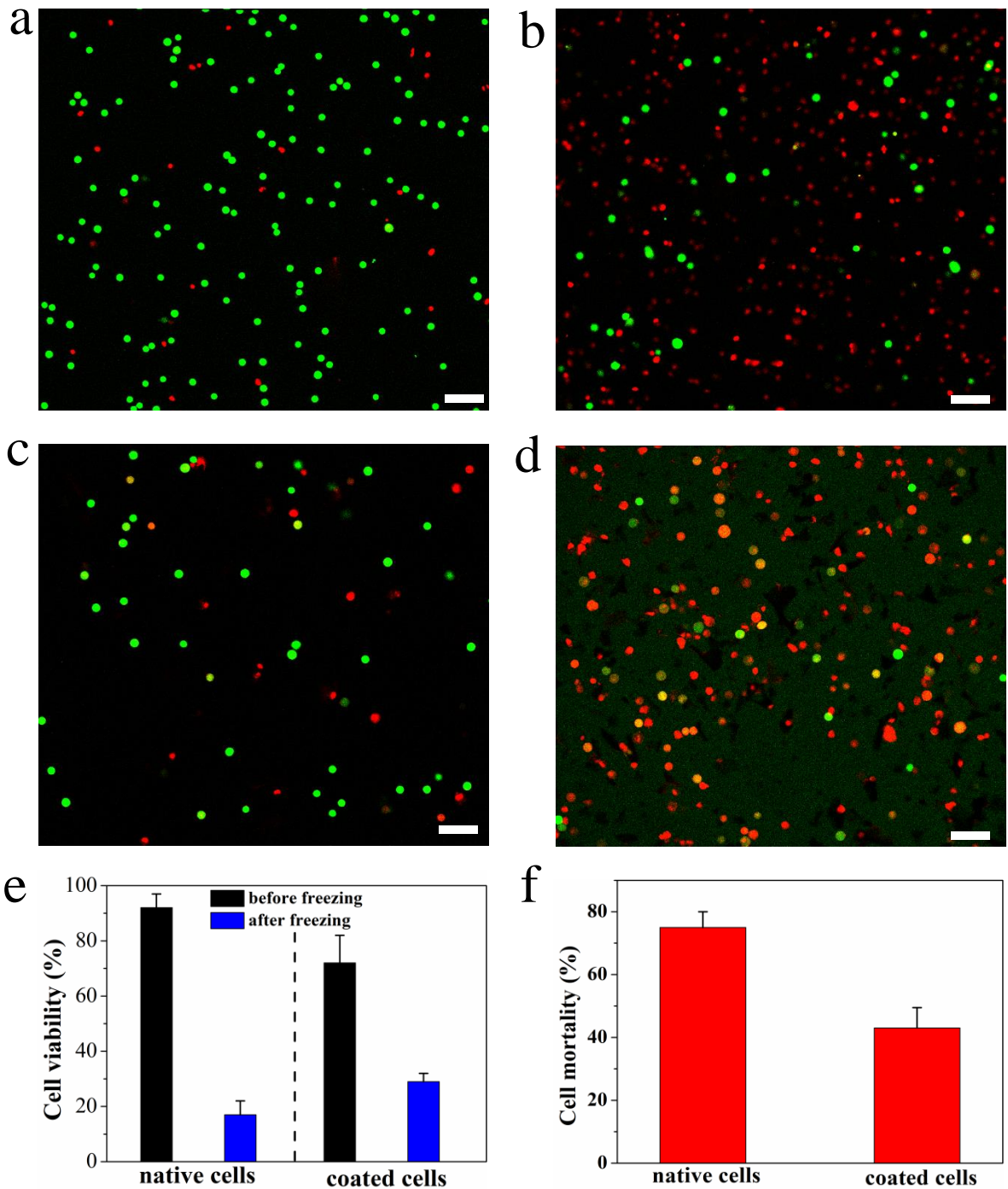

**Figure 7** The live/dead fluorescence images of native cells (a-b) and nanoshell coated cells (c-d). (a,c) Pre cryopreservation, and (b,d) post cryopreservation. Mammalian cells@[TA–$Fe^{3+}$]$_1$ were treated with FDA and propidium iodide (PI) for live/dead staining (green: live, red: dead). (e-f) Cell viability and mortality statistics before and after freezing and thawing. The scale bars are 100 μm.

## 4. Conclusions

In the present work, we modified the cell surface with different nanoshells in order to protect them from the mechanical damage of ice during cryopreservation. Directional thawing of coated cells was achieved and the cell viability was improved by the protection of premelting liquid layers and nanoshells. We elucidated the interaction mechanism between coated cells and the ice by the premelting theory and thermal regelation. The strength of interactions between cells and ice and the effective Hamaker constant were modified after the coating, which also increased thicknesses of premelted layers between cells and ice. Therefore, the enhanced liquid flow drove coated cells to the warm side in a thermal gradient and the directional migration was accelerating with time, while native cells stood still in the ice. Moreover, the increased premelted layers and the shells surrounding coated cells decreased cell mortality after freezing and thawing. This premelting-directed interface design strategy provides an effective way to modulate interactions between cells and ice, which is expected to further advance the cryopreservation of biological objects in various fields.

## Supporting Information

Supporting Information is available in the website.

## Acknowledgements

J.Y. acknowledges supports from National Natural Science Foundation of China (No. 51901190) and China Postdoctoral Science Foundation (2019M653728). L.S. acknowledges supports from the Fundamental Research Funds for the Central University (3102018jcc037, 3102019GHJD001).

**The table of contents entry**

**Interfacial interactions between ice and cells** were modulated by nanoshells via the premelting theory. The interaction mechanisms between coated cells and ice were analyzed.

**Keywords** freezing, interfacial design, nanoshells, cell/ice interactions, premelting

J. You, Y. Zhang, Q. Chen, Y. Xie, Y. Li, L. Hu, L. Shang*

**Title** Premelting-directed interfacial design of nanoshell-coated cells in directional freezing and thawing

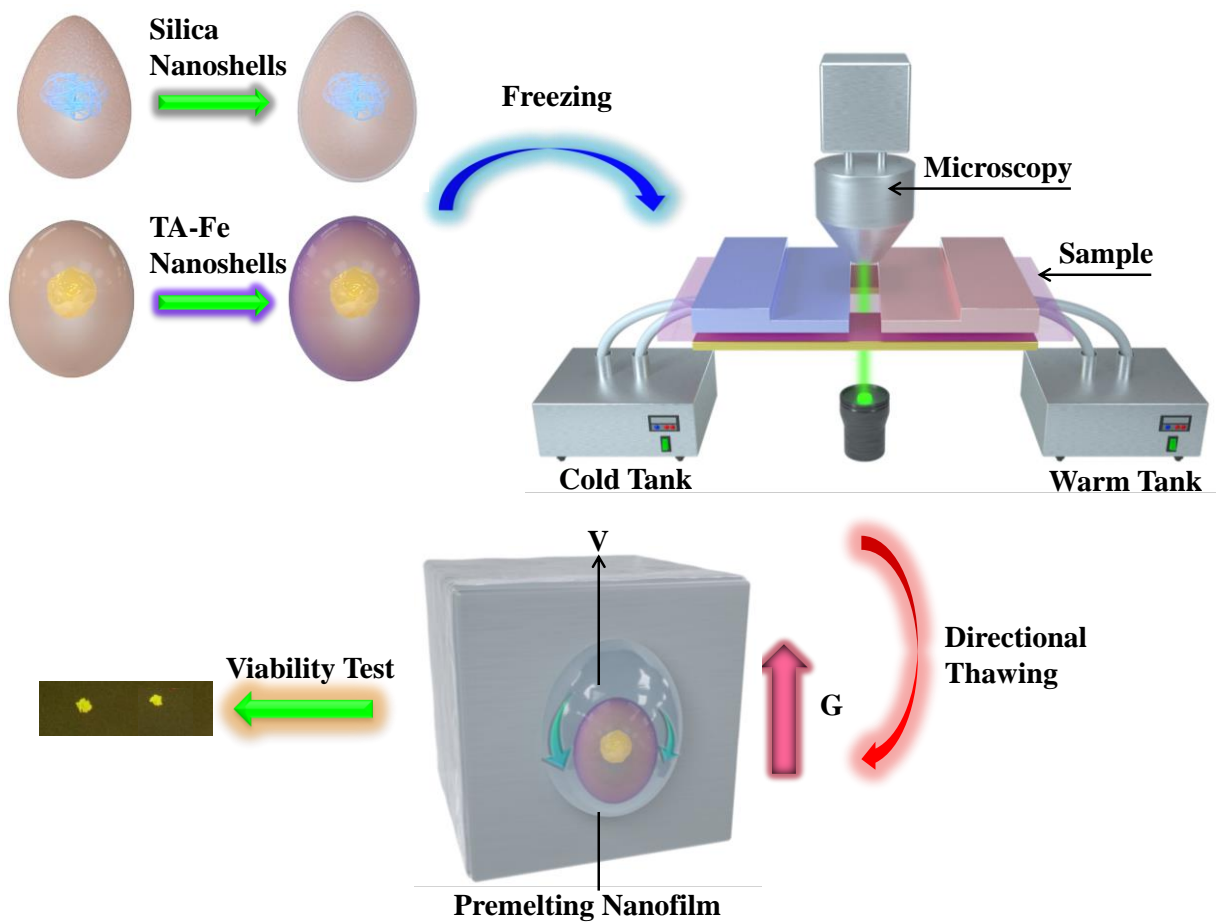

# Supporting Information

**Premelting-directed interfacial design of nanoshell-coated cells in directional freezing and thawing**

*Jiaxue You[1,2], Yunhan Zhang[1], Qiao Chen[1], Yifan Xie[1] Yixiao Li[1], Lifang Hu[3] and Li Shang[1,4]\**

[1] State Key Laboratory of Solidification Processing, Center for Nano Energy Materials, School of Materials Science and Engineering, Northwestern Polytechnical University, Xi'an, 710072, China

[2] Key Laboratory of Applied Surface and Colloid Chemistry, National Ministry of Education, Shaanxi Key Laboratory for Advanced Energy Devices, Shaanxi Engineering Lab for Advanced Energy Technology, Institute for Advanced Energy Materials, School of Materials Science and Engineering, Shaanxi Normal University, Xi'an 710119, China

[3] Key Laboratory for Space Bioscience and Biotechnology, Institute of Special Environmental Biophysics, School of Life Sciences, Northwestern Polytechnical University, Xi'an 710072, China

[4] NPU-QMUL Joint Research Institute of Advanced Materials and Structures (JRI-AMAS), Northwestern Polytechnical University,
Xi'an, 710072, China
\*Corresponding author: Li Shang, E-mail: li.shang@nwpu.edu.cn

**Contents**

Figures S1-S5

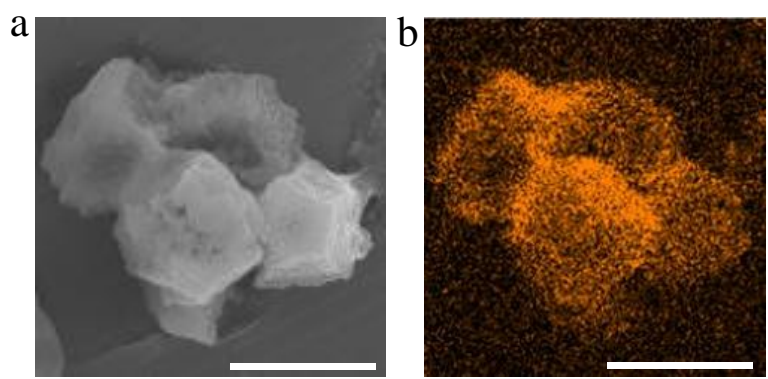

**Figure S1** SEM images and elementary mapping of yeast@$SiO_2$. (a) SEM of a coated yeast cell. (b) EDS spectroscopy plane profiles for Si. The scale bars are 5 μm.

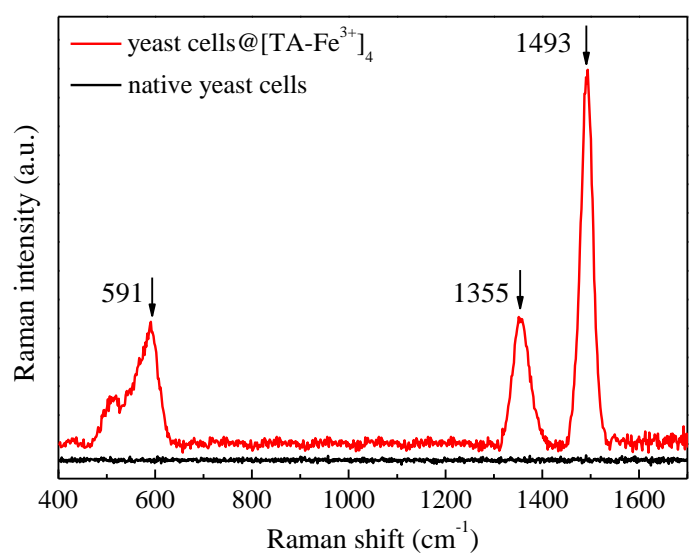

**Figure S2** Raman spectroscopy analysis of yeast cells@$[TA-Fe^{3+}]_4$.

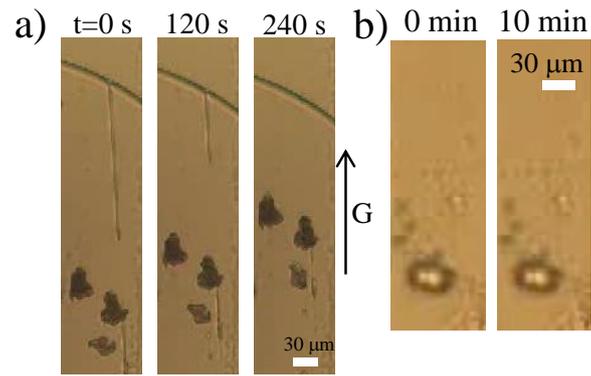

**Figure S3** In-situ observations of thermal regelation of bone cells@[TA-Fe$^{3+}$] and native cells: (a) coated cells: strong migration in the ice; (b) native cells: very weak migration in the ice.

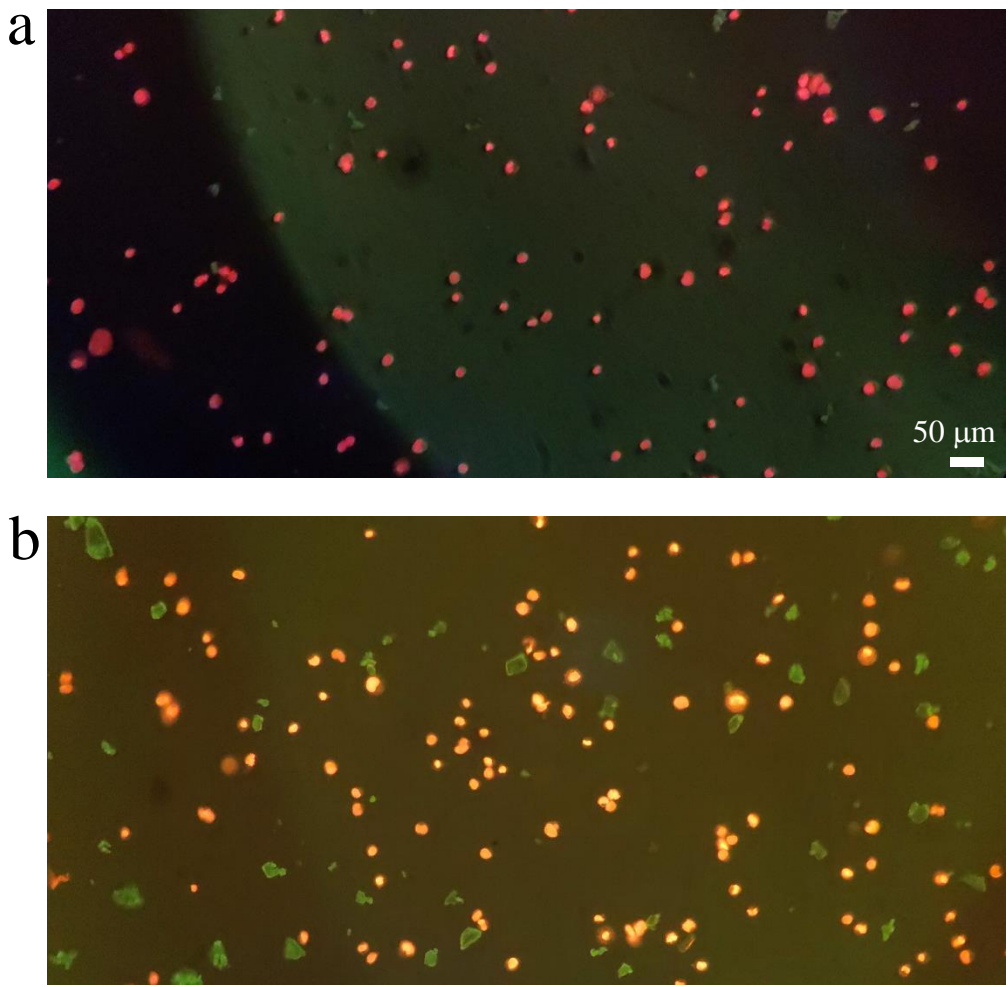

**Figure S4** The live/dead fluorescence images of native cells (a) and shell coated cells (b). Bone cells@[TA–$Fe^{3+}$]$_4$ were treated with FDA and propidium iodide (PI) for live/dead staining (green: live, red: dead) after freezing 72 hours. After freezing 72 hours, the coated cells have a higher viability (20%) than that of native cells (10%).

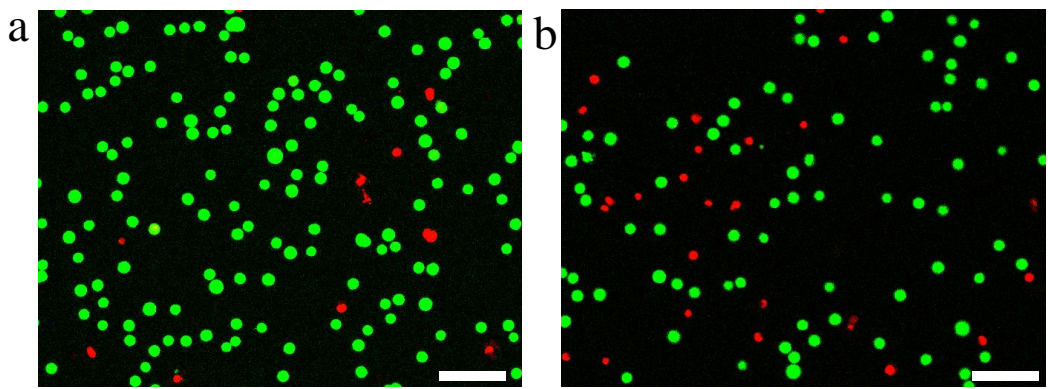

**Figure S5** The live/dead fluorescence images of native cells (a) and freezing cells with DMSO for 14 hours (b). The scale bars are 100 μm. After freezing 14 hours, the viability of freezing cells with DMSO is 79%.

27